# Charge transfer mechanism on $MoS_2$ nanosheets in the presence of a semiconductor photoactive media

Srinivasa Rao Konda,[1,*] Puspendu Barik,[1] Subshash Singh,[2] Venkatesh Mottamchetty,[1,3] Amit Srivasthava,[4] Rashid A. Ganeev,[5,6] Soma Venugopal Rao,[7] Chunlei Guo,[2] and Wei Li [1,*]

[1] The GPL Photonics Laboratory, State Key Laboratory of Luminescence and Applications, Changchun Institute of Optics, Fine Mechanics and Physics, Chinese Academy of Sciences, Changchun, Jilin 130033, China

[2] The Institute of Optics, University of Rochester, Rochester, NY, 14627, USA

[3] Department of Materials Science and Engineering, Uppsala University, Box 35, SE-75103 Uppsala, Sweden

[4] Department of Physics, TDPG College, VBS Purvanchal University, Jaunpur, 222001, India

[5] Institute of Theoretical Physics, National University of Uzbekistan, Tashkent 100174, Uzbekistan

[6] Department of Optics and Spectroscopy, Voronezh State University, 1 University Square, Voronezh 394006, Russia

7 School of Physics & ACRHEM (DIA-CoE), University of Hyderabad, Hyderabad 500046, Telangana, India

Correspondence: *ksrao@ciomp.ac.cn, *weili1@ciomp.ac.cn

**Abstract**

The studies of the nonlinear optical (NLO) properties of the transition metal dichalcogenides (TMDs) coupled with photoactive particles, plasmonic nanocavities, waveguides, and metamaterials remain in their infancy. This study investigates the third-order NLO properties of $MoS_2$ nanosheets in the presence of a semiconductor photoactive medium. Our extensive studies and the obtained results reveal the counteractive coupling effect of bare and passivated quantum dots on the $MoS_2$ nanosheet, as made evident by the analysis of the NLO processes. The enhanced NLO properties of $MoS_2$ nanosheets functionalized with CdSe and CdSe/$V_2O_5$ quantum dots are helpful for applications as saturable absorbers in laser applications and the emission of coherent short-wavelength radiation. The multiphoton-excitation resonance energy transfer mechanism exploiting remote dipole–dipole coupling, and ultrafast charge transfer pathways emerges as another plausible way to alter the NLO properties in TMDs.

Nonlinear optics emphasizes the interaction of intense light with matter and is also a fundamental building block of modern optics. Comprehensive understanding of ultrafast nonlinear optical (NLO) properties [polarization, $P(t) = \varepsilon\chi^{(n)}E^n(t), n \geq 2$], manifested in two-photon absorption (TPA), saturable absorption (SA), reverse saturable absorption (RSA), parametric frequency conversion, four-wave mixing, self-phase modulation, self-focusing, and optical Kerr effects, is the prerequisite for the development of many classical and quantum technologies. The transition metal dichalcogenides (TMDs), especially 2D TMDs, have demonstrated fascinating optical properties compared with graphene because of a sandwich-like structure of a transition metal atom layer between two layers of chalcogen atom, the optical band gap at room temperature, and thermodynamically stable structural phases, i.e., either trigonal prismatic (2H) or octahedral (1T) coordination of the metal atoms.[1–3] Among various TMDs, molybdenum disulfide ($MoS_2$), possessing more extensive interlayer spacing, exhibits outstanding NLO properties and potential applications as saturable absorbers and optical limiters due to their enhanced electronic correlations and significant transition dipole matrix elements.[2,4] Several researchers made various efforts to enhance the optical nonlinearity of 2D TMDs by exploiting the electric field, doping or strain, crystalline phase, presence of plasmonic hot carriers, and imposing multiple unique configurations like plasmonic nanocavities, waveguides, and metamaterials.[1,5–11] One reasonable way is to transfer energy or charge from adjacent photoactive particles such as luminescent quantum dots (QDs) and nanoparticles to obtain homogeneous enhancement of NLO responses of TMDs.[2,12,13] Therefore, it is crucial to find complete information to realize the controlling factors that affect the third-order NLO enhancement of the $MoS_2$ nanosheets and their mechanism in the presence of external factors under variable laser power.

QDs are potential candidates as a photoactive medium for the photo-excited energy or charge transfer process due to their excellent optical properties, e.g., strong multiphoton nonlinear absorption and photoluminescence (PL) quantum yield, having long-range dipole-dipole electromagnetic interactions.[14,15] However, the manipulation of NLO response in the $MoS_2$ nanosheet with the assistance of QDs and passivated QDs has not yet been explored much.[2] The transition metal (Mo) and chalcogen (S) atoms are bonded with a covalent bond in a hexagonal order for monolayer $MoS_2$. At the same time, the bulk crystals or multilayers (like a few layers nanosheet) are formed by the van der Waals attraction force between two adjacent monolayers. According to bond-charge theory, the nonlinear susceptibility ($\chi$) in TMD materials depends on the asymmetric charge distribution and the interatomic interaction (e.g., the covalent bonding, producing excess charge in the bonding regime), i.e., the dynamics of bond charge.[16–18] Under high laser power irradiation, the $MoS_2$ nanosheet depicts layer-dependent NLO processes – SA, RSA, and TPA.[19,20] However, third-order NLO responses enhanced/suppressed by the strong interaction in a complex of 2D nanostructures and semiconductor QDs have seldom been demonstrated thus far. Research in the material synthesis-based approach for layered van-der-Waals materials to control optical nonlinearities will open the way to previously unexplored processes.

In this perspective, the comprehensive understanding of the role of the QDs in determining the NLO properties of 2D-0D hybrid systems is critical to realizing their potential for future applications. Concerning the recent development of 0D–2D hybrid, various band gap-engineered QDs and TMDs are available to develop a suitable hybrid that may promote NLO properties by altering interfacial interaction, charge transfer, or energy transfer, depending on the future application. In this article, we demonstrate CdSe (core) and $CdSe/V_2O_5$ (core/shell, termed as V-CdSe now on) QDs embedded on the few-layer $MoS_2$ nanosheet to study comprehensively third-order NLO properties in exquisite detail. In our experiments, the hydrothermal technique was used to synthesize CdSe and V-CdSe QDs embedded on $MoS_2$ nanosheets (see the experimental section in supplementary information for details).[21,22] We investigate the effect of bare CdSe QD and passivated core/shell QD (V-CdSe) on third-order NLO processes. In this respect, the experimental observation and understanding of counteractive effects of CdSe and V-CdSe NLO processes using $MoS_2$ nanosheets is of fundamental interest and can lead to the knowledge of the energy and charge transfer mechanism responsible for the enhanced NLO performance.

# Results and discussion

## Charge Transfer mechanism

$MoS_2$ nanosheets can accommodate a large population of QDs due to a substantial surface-to-volume ratio and provide high in-plane carrier mobility. In the present 0D–2D hybrid systems, both QD (CdSe) and TMD ($MoS_2$ nanosheets) components have band gap alignment with a particular overlap architecture so that the system may promote interfacial interaction, charge transfer, or energy transfer. $MoS_2$/CdSe forms approximately type-II band alignment. The $CdSe/V_2O_5$ QDs can be categorized as staggered type II band alignment, where electrons or holes delocalize depending on their conduction band or valence band offset. Figure 1 shows the band offset for $MoS_2$, CdSe, and $V_2O_5$ using density functional theory calculations.[23–25] The bulk $MoS_2$, CdSe, and $V_2O_5$ bandgaps are 1.2 eV, 1.59 eV, and 2.73 eV, respectively.[23–25] In the present experiment, the size of the CdSe core was ~3 nm, and considering the diameter of the $CdSe/V_2O_5$ QDs as 4.5 nm, the thickness of the $V_2O_5$ shell is ~0.75 nm. Therefore, one can estimate the lowest excitonic bandgap will be at ~2.45 eV for core CdSe QDs.[26]

The excitation with photons of 800 nm (1.55 eV) and 400 nm (3.10 eV) has sufficient energy for exciton generation in the 0D-QDs and 2D-$MoS_2$ nanostructures. Besides, a hybrid exciton may form due to the electron transfer (ET) from the CdSe QDs to the $MoS_2$ nanosheets or via hole transfer (HT) in the opposite direction. Therefore, we expect photoinduced charge (electron) transport from photoexcited QDs to layered $MoS_2$ nanosheets, which affects nanosheets' NLO properties, described in the next section. The formation of the inorganic shell ($V_2O_5$) of core/shell structured CdSe QDs suppresses the surface trap states, thus improving surface passivation by preventing the interaction between the core and the environment. Numerous reports have confirmed that the shell passivation increases the quantum yield of the QDs and reduces the defect states, increasing the radiative lifetime, as the non-radiative rates decrease due to the Auger decay mitigation.[27] The $V_2O_5$ shell acts as a tunneling barrier for the photo-generated electron moving from the CdSe core onto the layered $MoS_2$ nanosheets, thus slowing down the ET process. In general, the ET (or HT) depends on the electron (or hole) effective mass in shell materials and the core/shell band offset (potential barrier).
The theoretical framework for charge transfer (CT) in weak donor–acceptor electronic coupling regime, developed by Marcus *et al.,* has been applied to describe charge transfer mechanisms in various molecular and biological systems.[28,29] It can explain the photoinduced electron transfer between QDs and layered $MoS_2$ nanosheets. According to Marcus model, the nonadiabatic rate of charge transfer ($k_{CT}$) depends on the driving force (i.e., Gibbs free energy, which depends on the difference between the conduction band levels of the donor and acceptor components, $-\Delta G°$), the nuclear reorganization energy for the charge transfer reaction ($\lambda r$) and the electronic donor-acceptor coupling strength ($V$).[30]

$$k_{CT} = \frac{2\pi|V|^2}{\hbar\sqrt{4\pi\lambda r k_B T}} \cdot exp\left(-\frac{(\lambda r + \Delta G^o)^2}{4\lambda r k_B T}\right) \tag{1}$$

Where, $k_B$ is a Boltzmann constant, and T is a temperature. Here, the solvent reorganization energy accompanying a CT process can be neglected using the dielectric continuum model calculation, producing negligible solvent reorganization energy.[31] $MoS_2$ nanosheet can be treated as multiple independent monolayers stacked together, and hence, $k_{CT}$ depends on the total electron-transfer rates of the QD with each $MoS_2$ monolayer. However, the values of $\Delta G°$ and V depend on the separation distance between QD and $MoS_2$ monolayers due to the change in the onset position of the conduction band as bandgap decreases from monolayer to multilayer conversion (direct to indirect transition happens). For $MoS_2$ nanosheets, i.e., multilayer $MoS_2$ in comparison to monolayer $MoS_2$, $-\Delta G°$ increases due to enhancement in the donor–acceptor energy band offset, and hence, the rate of electron transfer also increases from QDs to $MoS_2$, resulting in saturation of electron-transfer rate.[30] In a type-II band alignment like $MoS_2$-CdSe, photoexcited electrons and holes should transfer to opposite sides. They are expected to minimize their energy, leading to electrons bleaching the $MoS_2$ conduction band edge.

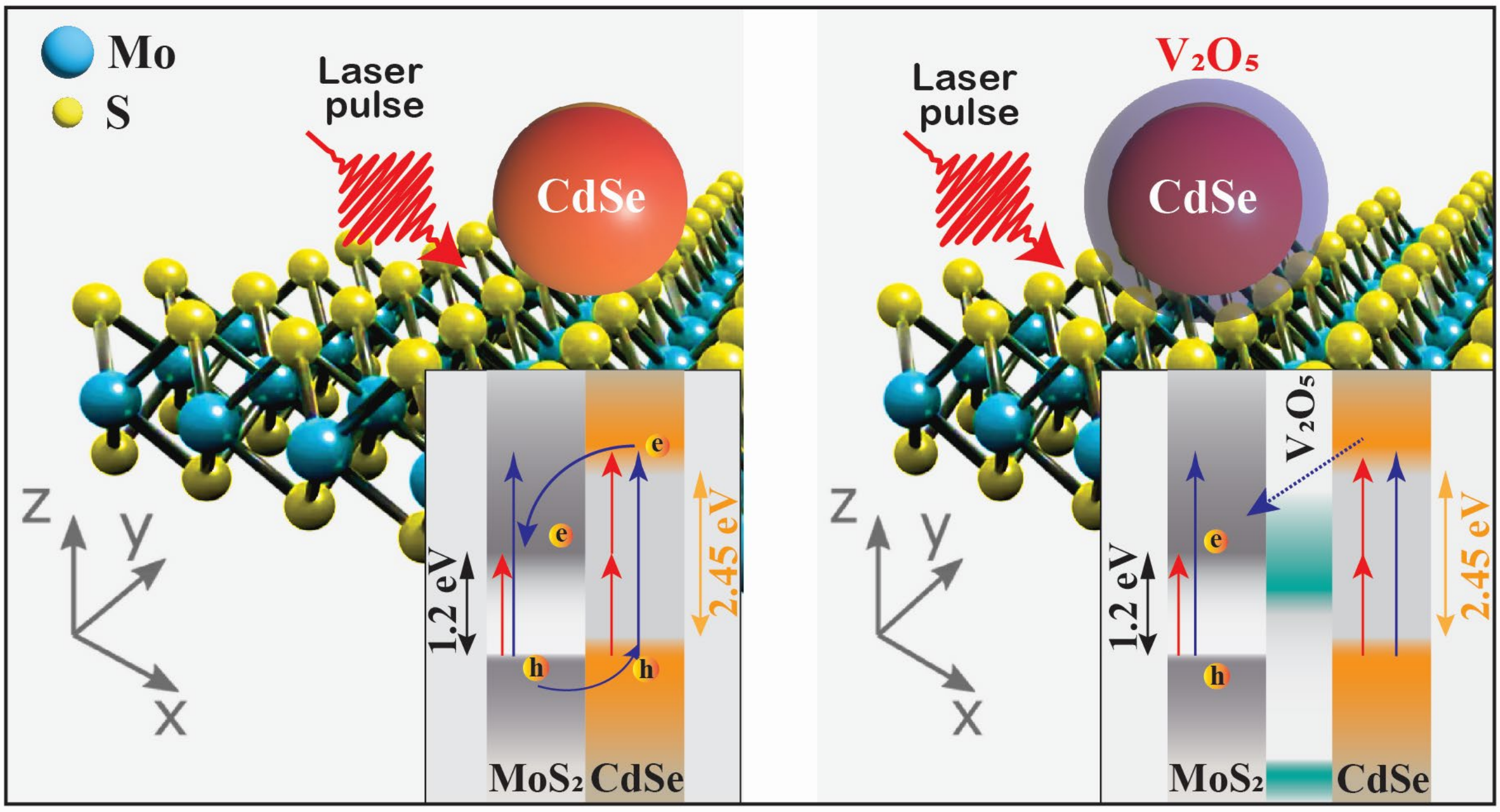

**Figure 1 Concept illustration of 0D−2D CdSe (CdSe/$V_2O_5$) QD−$MoS_2$ hybrids depicting band gap alignment of the acceptor ($MoS_2$) and donor (QD). The $V_2O_5$ shell acts as a tunneling barrier for the photo-generated electron moving**

**from the CdSe core onto the layered $MoS_2$ nanosheets, thus slowing down the electron transfer (ET) process. The interfacial trap states are not shown here. However, electrons are rapidly trapped at interfacial states at low fluence. The trap states are saturated at higher fluence, and long-lived electrons occupy the $MoS_2$ conduction band edge.**

Another important CT mechanism is Auger-assisted ET, where Auger recombination of an exciton becomes more efficient for strong exciton-exciton coupling at high excitation intensity. The effect is strong in low-dimensional systems due to the quantum and dielectric confinement effects, enhancing Coulomb coupling between the charges. In general, charge transfer dominates at a short distance of a few nanometers, while energy transfer dominates at a slightly longer distance. The primary energy transfer route is Förster-type non-radiative energy transfer (FRET), which involves dipole–dipole Coulomb interactions between a donor and an acceptor material, i.e., center-to-center distance. However, multiphoton-excitation resonance energy transfer may contribute in the NLO properties of $MoS_2$ nanosheets in presence of QD, where $MoS_2$ directly gains energy from the TPA dipole with strong oscillator strength in the QDs via remote Coulombic coupling.[2,30] The energy-transfer rate depends on energy overlap, i.e., an overlap between the donor (QDs) emission spectrum and the acceptor ($MoS_2$) absorption spectrum, and the Coulombic coupling strength, i.e., quantum mechanical oscillator transition densities of the donor polarization and acceptor polarization. An increased number of free carriers can be generated via photoexcitation for higher pump fluences beyond bandgap photoexcitation. Therefore, CT and energy transfer processes can impact the ultrafast third-order NLO response in $MoS_2$ nanosheet in the presence of core and core/shell QDs. The QDs can manipulate the third-order NLO properties of $MoS_2$ nanosheets via two possible explanations: (1) strong dielectric coupling between QD and nanosheets to modulate the nonlinear susceptibility or (2) strong electronic-vibrational coupling at the QD-$MoS_2$ interface. The ultrafast charge transfer in the $MoS_2$-CdSe improved the SA effect due to enhanced exciton coupling, and the shell passivation in $MoS_2$-V-CdSe reduced the NLO effect.

The $MoS_2$ nanosheets were synthesized through a hydrothermal approach in a Teflon-based autoclave at 210 °C, adapted from a method reported earlier.[22] CdSe (~3 nm, core) and CdSe/$V_2O_5$ (~4.5 nm, core/shell) QDs were synthesized using the hydrothermal method, described in detail in previously reported work.[21] The nanosheet-QDs composite structures were synthesized via the hydrothermal method by adding QDs synthesized previously, as briefly described in the Supporting Information (SI).[21,22]

## Morphology and Structural Characterizations

Figure 1a-b depicts the high-resolution transmission electron microscopy (HRTEM) image, affirming the presence of CdSe and CdSe/ $V_2O_5$ core/shell QDs on the surface of $MoS_2$ nanosheet. X-ray diffraction (XRD) was employed to study the crystal structure of three samples ($MoS_2$, $MoS_2$-CdSe, and $MoS_2$-V-CdSe), as depicted in Figure 2c. The XRD pattern of $MoS_2$ nanosheets shows the characteristic peaks much broader than those of bulk 2H-$MoS_2$ phase (JCPDS no.-00-037-1492, space group - P63/mmc), and no crystalline impurities were observed. The effective sizes of crystalline domain $MoS_2$ nanosheets are small, and the most intense diffraction peaks are not sharp but broad, arising from their significant disorder due to this buckling.[32] However, the nanosheets exhibit a significantly preferred orientation in (002) direction. $MoS_2$-CdSe shows diffraction peaks, matched with 2H-$MoS_2$ phase (similar to $MoS_2$ nanosheets) and wurtzite structured phase of CdSe (JCPDS no.-01-077-2307 hexagonal, space group - P63mc). However, low-intensity diffraction peaks are observed for $MoS_2$-V-CdSe (as indicated by the asterisk in Figure 2c), which are well matched with the orthorhombic phase of $V_2O_5$ (JCPDS no.-00-041-1426, space group - Pmmn), in addition to peaks arising from 2H-$MoS_2$ and wurtzite-CdSe phases. The (002) peak of the wurtzite phase of CdSe becomes strong and sharp due to $V_2O_5$ shell formation in $MoS_2$-V-CdSe (also (110) peak of the orthorhombic phase of $V_2O_5$ coincides with it) and the increase of CdSe QDs in composites. The noticeable decrease in peak broadening of the $MoS_2$-V-CdSe (4.5 nm) sample justifies the growth of the $V_2O_5$ shell on core CdSe as crystalline domain size increases further. The diffraction peaks corresponding to 2H-$MoS_2$ in $MoS_2$-CdSe and $MoS_2$-V-CdSe samples are slightly sharper and possessed less full-width half maxima (FWHM), which is an indication of the epitaxial growth of the shell formation on CdSe and proper coordination between $MoS_2$ nanosheet and QDs.[32,33]

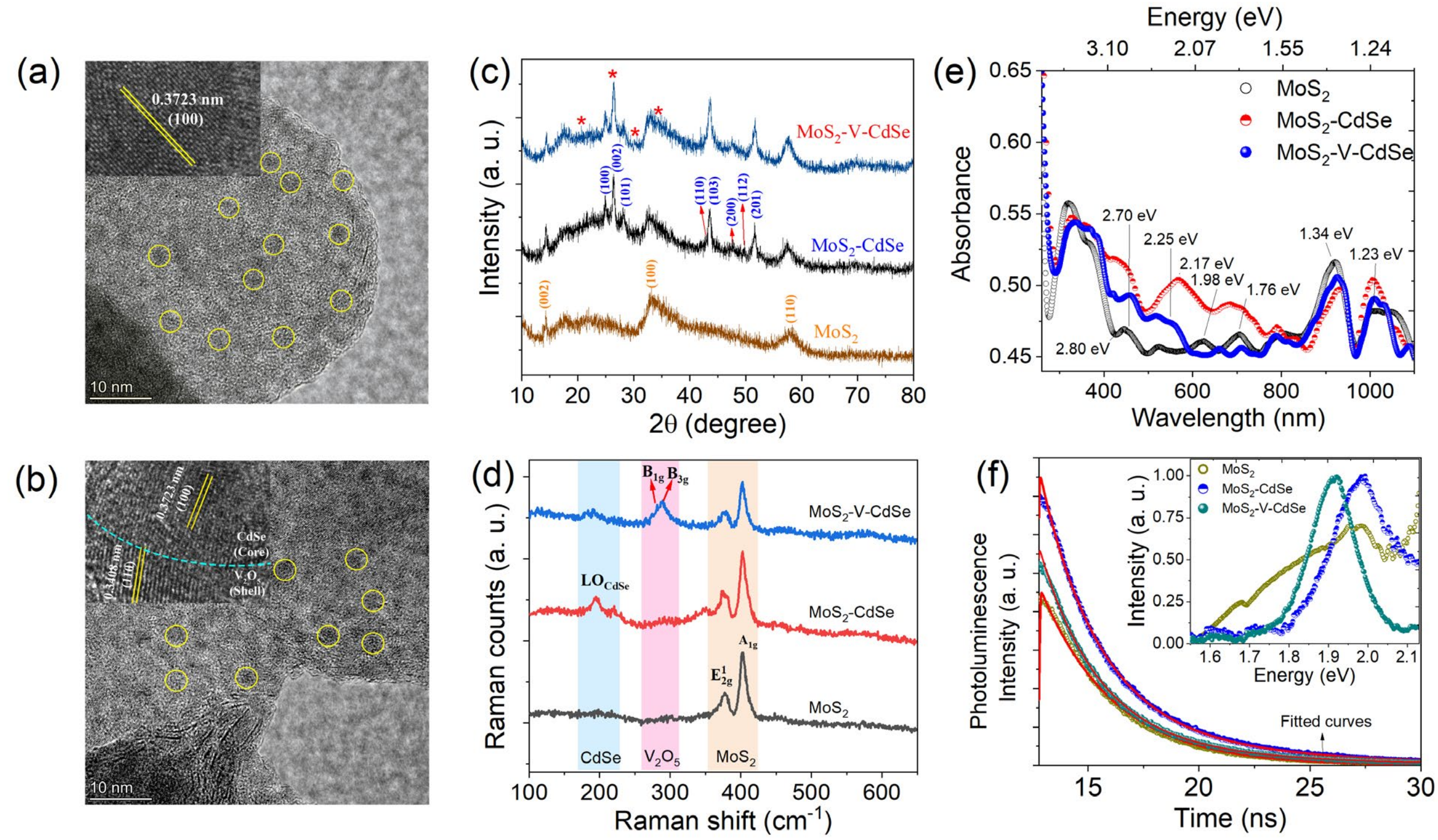

Figure 2 (a) Representative HRTEM image of *$MoS_2$-CdSe indicating CdSe QDs on the surface of $MoS_2$ nanosheet.* (b) The HRTEM image of *$MoS_2$-V-CdSe*. HRTEM image, lattice fringe image, high-angle annular dark-field (HAADF) image, and spectrum distribution with element ratio are provided in Figure S1 and Figure S2 (Supporting Information). (c) *XRD pattern of three samples - $MoS_2$, $MoS_2$-CdSe, and $MoS_2$-V-CdSe. Each (hkl) value in each curve represents the crystal planes (indicated by different colors), arising from CdSe, $V_2O_5$, and $MoS_2$ in various samples. The asterisk (*) represents low-intensity diffraction peaks at 20.3° (001), 26.1° (110), 32.4° (011), 34.3° (310) which are well matched with the orthorhombic phase of $V_2O_5$ (JCPDS no. 41-1426, space group - Pmmn). (d) Raman spectra of $MoS_2$, $MoS_2$-CdSe and $MoS_2$-V-CdSe. Color bars denote the positions of vibrational modes of different symmetry species arising from CdSe, $V_2O_5$, and $MoS_2$ in various samples.* (d) *Absorption spectra for $MoS_2$, $MoS_2$-CdSe and $MoS_2$-V-CdSe. (e) TRPL data and inset shows the PL spectra of $MoS_2$, $MoS_2$-CdSe and $MoS_2$-V-CdSe, excited at 532 nm (CW). Symbols represent the experimental data points while the solid lines are the theoretical fits.*

## Optical Properties and Band Structure

Figure 2d shows the representative Raman spectra where the characteristic in-plane ($E_{2g}^1$, at ~380.0 cm$^{-1}$) opposite vibration of two S atoms to the Mo atom and out-of-plane ($A_{1g}$, at ~402.5 cm$^{-1}$), vibrational modes of only S atoms in opposite directions confirm the high crystallinity of the $MoS_2$ nanosheets in all samples.[34,35] In off-resonance conditions, four first-order Raman active modes can be observed at 32 cm$^{-1}$ ($E_{2g}^2$), 286 cm$^{-1}$ ($E_{1g}$), 383 cm$^{-1}$ ($E_{2g}^1$) and 408 cm$^{-1}$ ($A_{1g}$) in bulk $MoS_2$.[35,36] The $A_{1g}$ mode is less affected by interlayer interactions but is sensitive to adsorbates on the $MoS_2$ surface.[35,36] However, it shows an increase in the peak width and redshift with the presence of CdSe and CdSe/$V_2O_5$ QDs, as shown in Figure 2d. The Raman peak corresponding to the longitudinal-optical (LO) mode for CdSe and V-CdSe is observed at 194 cm$^{-1}$ and 191 cm$^{-1}$, respectively, compared to bulk CdSe at 210 cm$^{-1}$. Among various vanadium pentoxide polymorphs, $\alpha$-$V_2O_5$ is thermodynamically stable at ambient conditions.[37] The orthorhombic crystal lattice of the $\alpha$-$V_2O_5$ consists of stacked layers of infinite $V_2O_5$ chains via ladder steps by weak interlayer interactions, which form several polymorph structures and have structural flexibility.[37,38] Ladder-like distortions dominate the $B_{3g}$ line at 291 cm$^{-1}$ and $B_{1g}$ line at 290 cm$^{-1}$, as shown in Figure 2d.[37]

The absorption spectrum of the few-layer $MoS_2$ nanosheet, $MoS_2$-CdSe, and $MoS_2$-V-CdSe suspension in water is shown in Figure 2e, where the Mie scattering-induced background was subtracted for clarity.

The monolayer $MoS_2$ is a direct semiconductor with a band gap of ~1.9 eV, while the multilayer or bulk $MoS_2$ is an indirect semiconductor with a narrower band gap of ~1.2 eV.[39,40] According to the GW approximation, the electronic bandgap of a freestanding monolayer $MoS_2$ has been predicted to be ~1.8 eV and decreased to 1.2 eV as the thickness increases or for the bulk.[40] The four characteristic absorption peaks of 2.80, 2.70, 1.98, and 1.76 eV (also termed as D, C, B, and A transitions, respectively) in the regions of 400–450 nm and 600–700 nm are the general features of 2H polytype $MoS_2$ nanosheets.[3,41,42] The absorption peaks, A and B, originate from the interband excitonic transitions at the K point of the 2D Brillouin zone of $MoS_2$. At the same time, their separation signifies the spin-orbit splitting of transitions at K.[41,42] On the other hand, the absorption peaks C and D are owing to the transitions between the higher density of state regions. Cadmium chalcogenide QDs, e.g., CdSe, are direct bandgap semiconductors with bulk bandgaps 1.74 eV with n-type semiconducting properties. For the zinc blend structure of CdSe QDs, the valance band splits into three distinct hole levels $1s_{h(A)}$, $1s_{h(B)}$, and $1p_h$, due to spin–orbit coupling according to effective mass approximation theory. The exciton transitions at 2.55 eV (485 nm), 2.25 eV (550 nm), and 2.13 eV (580 nm) correspond to a ($1p_h - 1p_e$) transition, ($1s_{h(B)} - 1s_e$) and ($1s_{h(A)} - 1s_e$) transitions, respectively.[43] The observed transitions also corroborate the theoretical approximation for $MoS_2$-CdSe and $MoS_2$-V-CdSe, though the contribution of $MoS_2$ prevails due to higher volume fraction in samples. The corresponding photoluminescence (PL) spectra are given in inset of Figure 2f. A few layers of $MoS_2$ show a broad peak with a center at 1.965 eV (~631 nm), though the PL intensity was very low and 10x enlarged in the figure. However, the characteristic peaks of $MoS_2$ are shifted in the $MoS_2$-CdSe and $MoS_2$-V-CdSe samples to 1.987 eV (~624 nm) and 1.918 eV (~646 nm), respectively, which may be attributed to the fact that the emission of the $MoS_2$ was altered due to charge transfer mechanism by the CdSe and CdSe/$V_2O_5$ QDs.

The photoinduced charge carrier recombination kinetics of $MoS_2$ nanosheets in the presence of QDs was characterized by time-resolved photoluminescence (TRPL) spectroscopy, where the information about the radiative electron–hole recombination after absorption of a short light excitation with ps pulse. Figure 2f depicts the TRPL decay curves from three samples. The TRPL decays have been fitted with a sum of exponential decays as $I_{TRPL}(t) = \sum_{i=1}^{n} A_i e^{-t/\tau_i}$, where $A_i$ and $\tau_i$ are the amplitude and lifetime of $i^{th}$ component, respectively. The Fitting parameters were shown in Table S1. Here, the decay curve for the $MoS_2$ nanosheet is fitted with mono-exponential ($n = 1$), and the obtained lifetime was 3.2614 ns. For the $MoS_2$ nanosheet with photoactive QDs, the decay curves are perfectly fitted with a bi-exponential ($n = 2$) fit, which is usually assigned to short-lived trap-mediated recombination ($\tau_1$) and long-lived radiative recombination ($\tau_2$). $MoS_2$-CdSe and $MoS_2$-V-CdSe show a short-lived recombination time of 2.5608 ns and 2.6525 ns, respectively, indicating faster recombination than pure $MoS_2$ nanosheet. $MoS_2$-CdSe and $MoS_2$-V-CdSe show a long-lived radiative recombination time of 7.2966 ns and 7.3926 ns, respectively.

## Third-order NLO properties

Intense pulsed radiation induces intensity-dependent transmittance, i.e., nonlinear absorption (NLA), which can be classified into two distinct phenomena. The first one is a single-photon nonlinear process, i.e., the SA resulting from population bleaching of the ground state owing to an excited state population, the transmittance of which increases with increasing optical intensity. The second process arises from multiphoton processes (e.g., TPA, multi-photon absorption, and RSA), during which the transmittance decreases with increasing optical intensity. Two paradoxical responses - SA and RSA are important in saturable absorbers (optical switches) for generating a pulsed laser and optical limiters for sensor protection, respectively. Figure 3 elaborately describes the NLA mechanisms to understand the NLA process's pump energy and intensity dependence in the $MoS_2$ and $MoS_2$-QDs composite. The electronic transitions occurring in layered 2D nanostructures must be considered to understand better the physical processes responsible for the NLA response of the hexagonal phases of $MoS_2$ nanosheet and their composites with QDs. The relatively rough optical absorption profiles (see Figure 2e) suggest that the occurring interband optical transitions should be associated with multiple phenomena simultaneously, e.g., SA, RSA, TPA, and their combinations.

The third-order NLO properties of the $MoS_2$ nanosheets, $MoS_2$-CdSe and $MoS_2$-V-CdSe were investigated using an open-aperture (OA) and closed-aperture (CA) Z-scan measurements with femtosecond laser pulses at 800 nm ($E_{800\ nm}$= 100, 200, 280, and 400 nJ, corresponding peak intensities $I_{800\ nm}$= 67, 135, 188, and 269 GW/cm$^2$) and 400 nm ($E_{400\ nm}$= 30, 50, 100, and 280 nJ, corresponding peak intensities $I_{400\ nm}$=20, 33, 67 and 188 GW/cm$^2$). The pulse width of the laser source (Ti: Sapphire) at 800 nm was ~35 femtoseconds (fs) with a repetition rate of 1 kHz. In comparison, the 400 nm emission was produced during second harmonic generation using a 0.2 mm β-BBO Type-I crystal. The $MoS_2$ nanosheets exhibit significant SA for the fs pulses, resulting in the imaginary part of third-order susceptibility $(Im[\chi^{(3)}]) \sim\ 10^{-15}\ esu$, the figure of merit (FOM) = $|Im[\chi^{(3)}]/\alpha_0|\ \sim 10^{-15}\ esu\ cm$ and free-carrier absorption cross-section of $\sim\ 10^{-17} cm^2$.[44] The precision of the experimental setup was confirmed by our previous measurements of NiO/$WS_2$ (2D),[45] Ni-$CsPbBr_3$ (2D),[46] and Ag[47] nanoparticle suspensions. The normalized optical transmittance of $MoS_2$ nanosheets, $MoS_2$-CdSe and $MoS_2$-V-CdSe with respect to different laser intensities is shown in Figure 3a-c. We also fitted the experimental data with various NLA processes using eq. 2 to eq. 4. The transmittance peak (at Z = 0) becomes higher and broader with increasing pump intensity initially, indicating a strong SA effect, i.e., single-photon absorption because the excitation photon energy (1.55 eV) is higher than the bandgap ($E_g$) of $MoS_2$ nanosheet (1.2 eV) for three samples (Scheme I-III, Figure 3d). To extract the contribution coming from either CdSe QDs or passivated CdSe/$V_2O_5$ QDs, we determined the ratios of OA Z-scan data of $MoS_2$-CdSe/$MoS_2$, $MoS_2$-V-CdSe/$MoS_2$-CdSe for 800 nm (top panel) and 400 nm (bottom panel) at four different pumping energies, as shown in Figure S3. The data fit well with theoretical equations (eq. 2 to eq. 4, Method section). However, the contribution arising from CdSe QDs and passivated CdSe/$V_2O_5$ QDs are opposite except for two cases at 200 nJ (800 nm) and 50 nJ (400 nm).

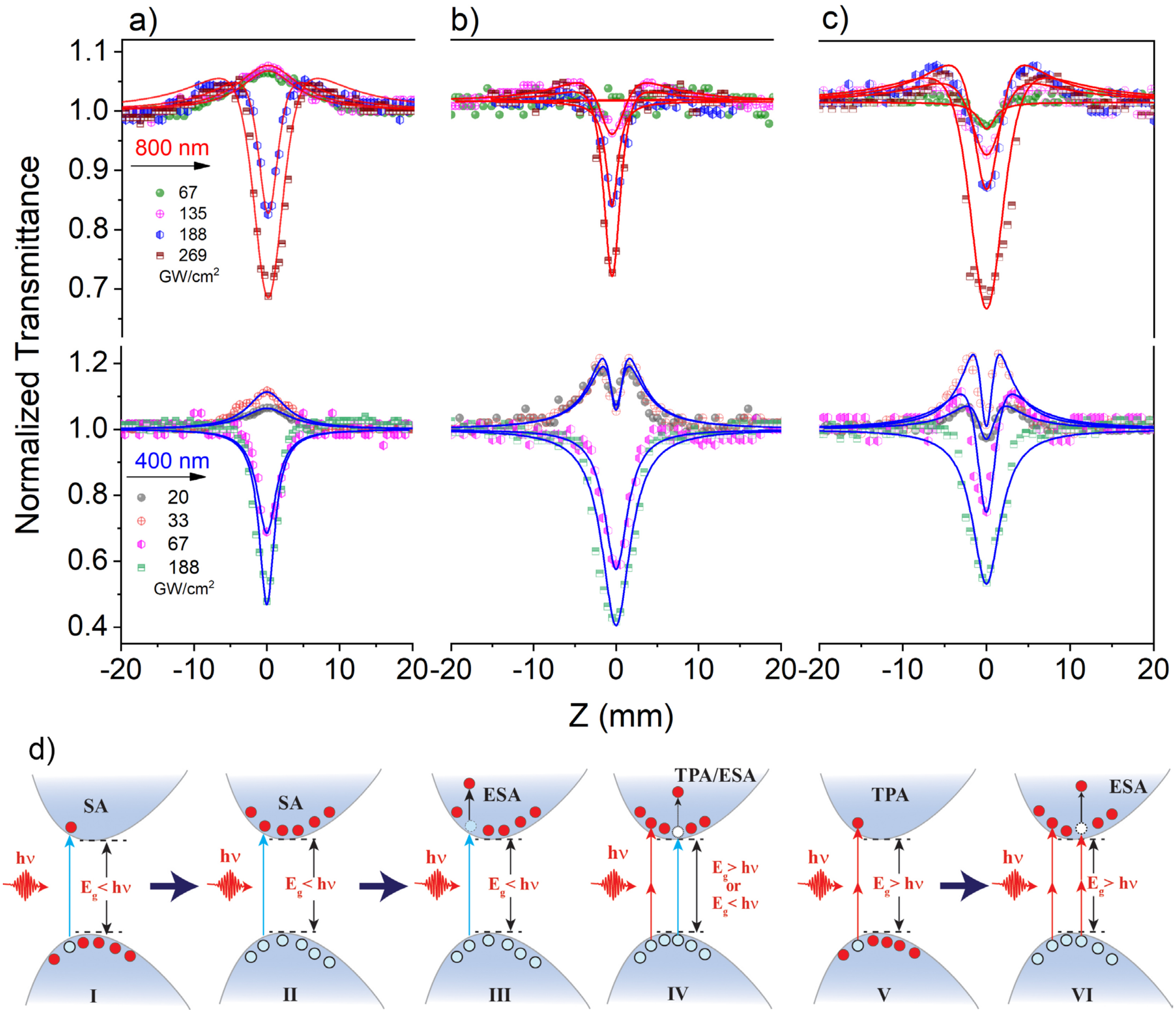

**Figure 3 Pump intensity-dependent OA Z-scan results of (a) $MoS_2$ nanosheets, (b) $MoS_2$-CdSe, and (c) $MoS_2$-V-CdSe for 800 nm (top panels) and 400 nm (bottom panels), respectively. Symbols denote the experimental data and solid lines**

**represent the fitted curves corresponding to the NLO processes described in the Method section (eq. 2 to eq. 4). (d) Schematics of the NLA mechanism under different pump energy (hν) and intensities. (I-III) the schemes show that the intensity increases gradually, and various NLA processes occur under other conditions. Initially, SA appears at low fluence with pump energy more significant than the bandgap (Eg). Further intensity increase induces SA and RSA processes. Electrons in the conduction band also absorb pump energy and promote higher energy levels. Scheme IV designates the RSA process (sometimes SA + RSA) at high fluence when exciting beam energy is higher or lower than the bandgap. Scheme V describes the TPA process at high fluence when the pump energy is less than the bandgap ($E_g$). Scheme VI describes the RSA process at a high fluence of pump pulse with energy hν, which is lower than the bandgap ($E_g$).**

Upon 400 nm (3.10 eV) and 800 nm (1.55 eV) irradiation, valence band electrons can be excited to high energy levels in the conduction band via different processes under sufficient intensity fs laser pulse (here, 20 – 269 GW/cm$^2$), followed by the conversion of the photo-generated electron−hole pairs to hot carriers. These hot carriers cool down shortly within the ps time scale through two different mechanisms – (i) carrier−carrier (~2 ps) and (ii) electron−phonon scattering (~34 ps) for the 2H-$MoS_2$.[11,48] Therefore, the photo-excited carriers generated by a fs laser excitation relax and populate in the valence and conduction bands, creating an equilibrium electron−hole distribution depending on the incident laser intensity. According to the Pauli-Blocking phenomenon (owing to the Pauli exclusion principle), at an intense laser intensity, the number of excited carriers accumulated in the conduction band is progressively increasing, and further transitions will be forbidden, resulting in a reduction of photon absorption after depletion of all the empty band states, as depicted in Figure 3d (Scheme II). Degenerate semiconductors exhibit a similar phenomenon, i.e., Burstein-Moss (BM) shift, in which electrons are pumped to higher energy states close to the conduction band edge from the valence band due to the bottom of the conduction band already being populated. The BM shift can be ruled out for our case. The OA Z-scans show the evidence in Figure S3. Pristine $MoS_2$ nanosheets and composite samples exhibit SA behavior at low laser intensities due to the Pauli blocking mechanism. However, the SA response turns to RSA response at higher laser intensity levels due to TPA processes, as shown in Figure 3d (Scheme III and VI).

Our observation for three samples shows that the pristine $MoS_2$ nanosheets possess SA at 100 and 200 nJ pulse energies of 800 nm emission. With further increase in the laser energy, i.e., at 280 and 400 nJ, $MoS_2$ nanosheets possess SA+RSA, which corroborates the previously reported NLA process for $MoS_2$.[4,30,49] In the case of $MoS_2$-CdSe at 800 nm, with 100 nJ pulses, no NLA properties were observed. This might be due to a conflict of SA from $MoS_2$ and the same amount of RSA contribution from CdSe QDs, as evident by the ratio curve in Figure S3. However, at 200 nJ, we observed the SA+RSA (Scheme II-III). Further increase in laser energy causes the prevalence of RSA at 280 and 400 nJ (Scheme – III). The valley's depth is increased with a further increase in laser energy (as shown in Figure 3a-c, top panel, $MoS_2$-V-CdSe panels at 200, 280, and 400 nJ).

The experimental observations can be well understood by the charge transfer mechanism described in the previous section, as shown in Figure 4. Since 1.55 eV photons are insufficient for CdSe-QD excitation, the SA can occur in $MoS_2$-CdSe as a consequence of the following mechanism: (a) energy transfer from the excited 0D-QDs to the 2D-$MoS_2$, (b) HT from the 2D-valance band maxima (VBM) to the 0D-VBM, or (c) ET from the 0D-conduction band minima (CBM) to the 2D CBM. Because the 0D-QD bandgap (here ~2.45 eV for CdSe QDs) is greater than that of the 2D-$MoS_2$ (~1.2 eV), energy transfer from the 2D-$MoS_2$ nanosheet to the 0D-QD is impossible. However, hole transfer from the 2D-$MoS_2$ VBM to 0D-VBM is also possible based on the band energy alignment shown in Figure 4a-b, where the 0D-VBM is higher than the 2D-VBM. The TPA increases for $MoS_2$-V-CdSe at 100 nJ probably due to the passivation of CdSe QDs by the $V_2O_5$ shell (now the ET process is somehow restricted), while NLA remains the same with an increase in laser energy at 200, 280, and 400 nJ, i.e., SA+RSA. The $V_2O_5$ shell creates interfacial states between $MoS_2$ and CdSe (Figure 4c). At low fluence, electrons trap at interfacial states, but they saturate the trap states at higher fluence. Consequently, it allows long-lived electron occupation of the $MoS_2$ conduction band edge, in which carrier trapping and electron transfer both occur.

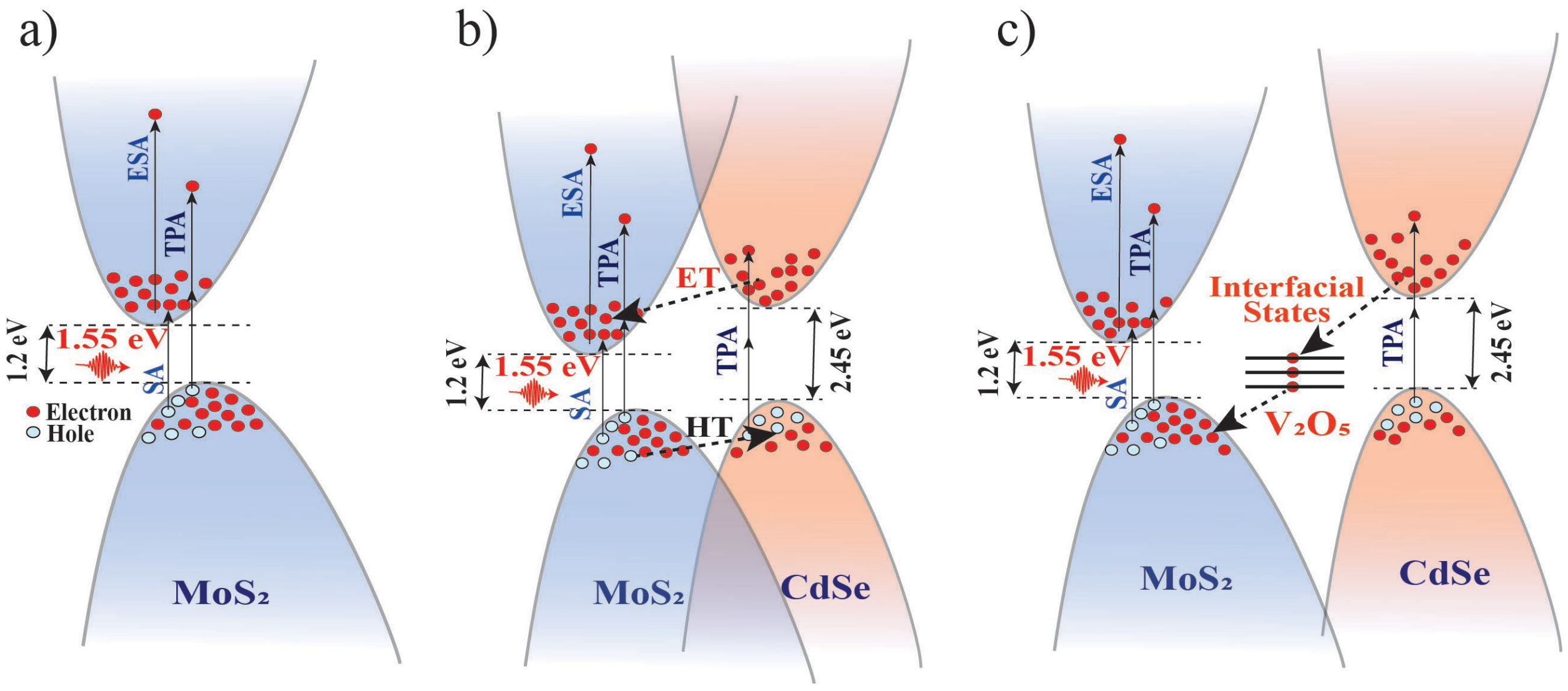

**Figure 4 Schematic charge transfer mechanism and energy-level diagram for (a) $MoS_2$, (b) $MoS_2$-CdSe, and (c) $MoS_2$-V-CdSe. However, the basic mechanisms are same as Figure 3 (scheme I – VI) depending on the pump energies and fluence.**

In the case of 400 nm OA Z-scans (Figure 3a-c, bottom panel), pristine $MoS_2$ nanosheets possess SA at 30 and 50 nJ of 400 nm, depicting the RSA with a further increase in the laser energy up to 100 and 280 nJ. The $MoS_2$-CdSe and $MoS_2$-V-CdSe show similar SA+RSA processes at 30 and 50 nJ; however, the RSA increases near the focal plane in the case of the $MoS_2$-V-CdSe due to the $V_2O_5$ shell contribution, i.e., saturation in the trap states at higher fluence, as shown in Figure 4c. The RSA for $MoS_2$-CdSe, compared to pristine $MoS_2$ nanosheet, shows opposite behavior at 100 and 280 nJ, i.e., the $MoS_2$-CdSe possess SA near high intensities, whereas $MoS_2$-V-CdSe shows SA+RSA at 100 nJ and RSA at 280 nJ. At low fluence, the ET mechanism is predominant for the $MoS_2$-CdSe (Figure 4).

Figures S4a and S4b show the CA Z-scan data for $MoS_2$, $MoS_2$-CdSe, and $MoS_2$-V-CdSe at different laser energies (like OA Z-Scans) at 800 nm and 400 nm fs pulses. All samples display the self-focusing nonlinear refraction (NLR) process with positive values of $\gamma$. However, there is a difference in the values of $\gamma$ due to the contribution of CdSe and passivated CdSe/$V_2O_5$ QDs. For shorter laser excitation pulses (*fs*), the most sizable contribution is from the faster response of the bound electronic and molecular reorientation nonlinearities prevailing over free carrier refraction. The NLR process originates from the redistribution and deflections of the bound charge, i.e., the electronic Kerr effect, inducing an intensity dependence of the NLO refraction. Consequently, in our study, thermal lensing, i.e., much slower processes (*ps* to*ns* regime) arising from the thermo-optic effect, does not contribute to NLR under excitation of 35 fs pulse (at 800 nm with a repetition rate of 1 kHz). Similar to OA data, Figure S4c shows the ratios of CA data to extract the contribution from QDs and passivated QDs. To quantitatively determine the values of the nonlinear absorption coefficient ($\beta$), nonlinear refraction index ($\gamma$), and saturation intensities ($I_{Sat}$), OA and CA Z-scan data were fitted by assuming the nonlinear absorption models equations for normalized transmittance, presented by eq. 2 to eq. 5 (see Method section for details). The third-order complex susceptibility $\chi^{(3)}$ was calculated using eq. 6 to eq. 8 (see Method section for details). In the present case, samples show various nonlinear processes like TPA, SA, RSA, the combined process of SA+RSA, and nonlinear refraction and absorption (NRA). We determined the third-order nonlinear susceptibilities ($\chi^{(3)}$) of the suspensions of three samples using the fs 800 nm and 400 nm pulses with four different energies. The values of $\beta$, $\gamma$, $I_{Sat}$, and $\chi^{(3)}$ are presented in Table S2. Therefore, the experimental observations suggest that the CdSe QDs and passivated CdSe/$V_2O_5$ QDs act differently on the NLA properties of $MoS_2$ nanosheets. The NLO parameters extracted from the ratio data by theoretically fitting with the NLO process at different excitation wavelengths are summarized in Table. S3.

# Method

## Characterization of samples

The morphology and components of the samples, including high-resolution transmission electron microscopy (HRTEM) image, lattice fringe image, high-angle annular dark-field imaging (HAADF), and spectrum distribution with element ratio were performed using a field emission TEM (JEM-2100F). X-ray diffraction (XRD) patterns were recorded by an X-ray diffractometer (Bruker D8, Germany), taking monochromatic Cu-$K_\alpha$ as a radiation source. The UV-Visible (UV-Vis) spectra were acquired with a spectrophotometer (Cary 5000 UV-Vis-NIR, Agilent) at room temperature (25℃) in the 300 to 1000 nm range with a scanning interval of 1 nm. Raman spectra were collected using LabRAM HR Evolution spectrometer (Jobin Yvon Horiba) equipped with a continuous-wave laser at 532 nm (spot size of ≈ 1 μm, average power at sample ~ 1 mW) focused by a microscope objective (100x, NA=0.9) and dispersed by 1800 lines $mm^{-1}$ grating. Photoluminescence spectra were obtained from a film fabricated on a Silicon (Si) substrate using the LabRAM HR Evolution spectrometer at room temperature in air (excited at 532 nm with a spot size of ≈ 1 μm, average power at sample ~ 100 μW) focused by a microscope objective (100x, NA=0.9) and dispersed by 600 lines $mm^{-1}$ grating. The self-built time-resolved fluorescence spectrometer testing platform were employed to measure the fluorescence lifetime of samples using a pulsed 532 nm Supercontinuum laser (OYSL Photonics, SC-Pro, 150 ps pulse lengths) with a 5 MHz repetition rate which was focused on the sample after an objective lens ($NA = 0.4$) with a power of 238 μW when focusing the sample. The detection system employed for this study consisted of an SPCM-AQRH single photon-counting module (SPCM-AQRH-15, Excelitas Technologies), and the lifetime module utilized was the TimeHarp 260 P.

## Z Scan

An open-aperture (OA) and closed aperture (CA) Z-scans were used to study the ultrafast NLO properties of the $MoS_2$ nanosheets, $MoS_2$-CdSe, and $MoS_2$-V-CdSe nanocomposites dispersions in water. Initially the powdered samples were dispersed in double distilled water and ultrasonication for 30 minutes, then after the aqueous solution was placed in a 1 mm thick quartz cuvette, and their transmittance was measured as a function of incident laser intensity. The optical arrangement was similar to that used in previous experiments[50] and a brief description is given in supplementary information. Z-scan technique is an important technique for characterizing the optical nonlinearity due to its high sensitivity and simplicity. The values of nonlinear absorption coefficient ($\beta$) and nonlinear refraction index ($\gamma$) can be determined using OA and CA Z-scan measurements, respectively. In the case TPA, SA, RSA and combined process of SA+RSA and NRA, the normalized transmittance (T) of laser beam can be defined by [51–55]

$$T_{TPA}(z) = T_{RSA}(z) \approx 1 - \frac{q(z)}{2\sqrt{2}} \tag{2}$$

$$T_{SA}(z) = 1 + \frac{I_0}{I_{sat}(x^2+1)} \tag{3}$$

$$T_{SA+RSA}(z) = \left(1 - \frac{q(z)}{2\sqrt{2}}\right) \times \left(1 + \frac{I_0}{I_{sat}(x^2+1)}\right) \tag{4}$$

$$T_{NRA}(z) = 1 + \frac{2(-\rho x^2 + 2x - 3\rho)\Delta\Phi_0}{(x^2+1)(x^2+9)} \tag{5}$$

where, $q(z) = I_0 \beta L_{eff}/(1 + z^2/z_0{}^2)$, $I_0$ is the laser peak intensity at focus ($z = 0$), $x = z/z_0$, $z_0 = kw_0^2/2$ is the Rayleigh length or range, $k = 2\pi/\lambda$ is the wave vector, $w_0$ is the radius of the beam waist, $\rho = \beta/2k\gamma$, and $\Delta\Phi_0 = k\gamma I_0 L_{eff}$ is the phase change due to nonlinear refraction at focus ($z = 0$). The effective length of the nonlinear medium can be expressed as, $L_{eff} = \left(1 - exp(-\alpha_0 L)\right)/\alpha_0$, where $\alpha_0$ is the linear absorption coefficient and $L$ is the thickness of the sample. We have to emphasize that the above formulas are only valid with a good quality Gaussian beam ($M^2 \approx 1$), and thin nonlinear samples ($L < z_0$). In our experiments, the 800 nm and 400 nm beam's spatial characteristics fulfill this requirement. The third-order complex susceptibility $\chi^{(3)}$ can be defined using the relation[52]

$$\chi^{(3)} = Re\left[\chi^{(3)}\right] + i \cdot Im\left[\chi^{(3)}\right] \tag{6}$$

The values of nonlinear absorption coefficient ($\beta$) can be deduced by OA Z-scan measurement. By defining $\beta$, the corresponding imaginary part of third-order susceptibility ($Im\left[\chi^{(3)}\right]$) can be calculated following the eq. 7.

$$Im\left[\chi^{(3)}\right] = \frac{c^2 n_0^2 \varepsilon_0 \beta}{\omega} \tag{7}$$

By using the nonlinear refraction coefficient ($\gamma$) measured from CA Z-scan measurements, the corresponding real part of third-order optical susceptibility ($Re\left[\chi^{(3)}\right]$) is obtained using the following relation eq. 8 :

$$Re\left[\chi^{(3)}\right] = 2cn_0^2 \varepsilon_0 \gamma \tag{8}$$

where, $n_0$ is the linear refractive index, $\varepsilon_0$ is the vacuum permittivity, and $\omega$ is the angular frequency of the laser beam. From eq. 7 and eq. 8, the magnitude of third-order susceptibility ($\chi^{(3)}$) is obtained as

$$\left|\chi^{(3)}\right| = \sqrt{Re[\chi^{(3)}]^2 + Im[\chi^{(3)}]^2} \tag{9}$$

Supporting Information

# Charge transfer mechanism on $MoS_2$ nanosheets in the presence of a semiconductor photoactive media

Srinivasa Rao Konda,[1,*] Puspendu Barik,[1] Subshash Singh,[2] Venkatesh Mottamchetty,[1,3] Amit Srivasthava,[4] Rashid A. Ganeev,[5,6] Soma Venugopal Rao,[7] Chunlei Guo, [2] and Wei Li [1,*]

[1] The GPL Photonics Laboratory, State Key Laboratory of Luminescence and Applications, Changchun Institute of Optics, Fine Mechanics and Physics, Chinese Academy of Sciences, Changchun, Jilin 130033, China
[2] The Institute of Optics, University of Rochester, Rochester, NY, 14627, USA
[3] Department of Materials Science and Engineering, Uppsala University, Box 35, SE-75103 Uppsala, Sweden
[4] Department of Physics, TDPG College, VBS Purvanchal University, Jaunpur, 222001, India
[6] Institute of Theoretical Physics, National University of Uzbekistan, Tashkent 100174, Uzbekistan
[7] Department of Optics and Spectroscopy, Voronezh State University, 1 University Square, Voronezh 394006, Russia
[8] School of Physics & ACRHEM (DIA-CoE), University of Hyderabad, Hyderabad 500046, Telangana, India

Correspondence: *ksrao@ciomp.ac.cn, *weili1@ciomp.ac.cn

## Chemicals

Sodium molybdate dihydrate ($Na_2MoO_4$. $2H_2O$), thiourea ($(NH_2)_2CS$), ethanol, (all purchased from Sigma Aldrich). Dodecylamine (DDA) (98%, Alfa Aesar), Vanadium (III) chloride (99%, Alfa Aesar), Toluene ($C_6H_5.CH_3$, Thomas Baker), and Methanol ($CH_3.OH$, Fisher Scientific). $Li_4[Cd_{10}Se_4(SPh)_{16}]$ clusters were prepared similarly as described previously.[1] All chemicals were reagent grade and used as received without further purification. High-purity deionized (DI) water was used for washing after synthesis.

## Synthesis of $MoS_2$ nanosheets

A solvothermal method was used to prepare a few layers of pure $MoS_2$ nanosheets.[2] In a typical synthesis, 1 mmol (242 mg) $Na_2MoO_4$. $2H_2O$ and 5 mmol (380 mg) thiourea were dissolved in 60 ml (30 ml ethanol + 30 ml DI water) under continuous stirring for 30 minutes at room temperature to form homogeneous solutions. Finally, the solution was put in a 100 ml Teflon-lined stainless-steel Autoclave at 210 °C for 24h. After cooling to room temperature, the sample was washed several times with a mixture of DI water and ethanol to remove impurities. Finally, the sample was dried in a vacuum oven at 80 °C overnight to obtain pure $MoS_2$ nanosheet.

## Synthesis of CdSe and CdSe/$V_2O_5$ core/shell QDs (denoted as V-CdSe)

CdSe QDs were synthesized by a method reported earlier.[3] In a typical synthesis, 120 mg of $Li_4[Cd_{10}Se_4(SPH)_{16}]$ cluster was added to 10 ml of DDA at 100 °C and then kept at 120 °C for 30 minutes. Next, the temperature was raised to 220 °C (10 °C/min) and maintained for nearly 3 hours. To synthesize CdSe/$V_2O_5$ core/shell QDs, $VCl_3$ (10 mg) was added at 120 °C. Then, the temperature was raised to 220 °C and maintained for 3 hours. The mixture was finally allowed to cool to the room temperature in the nitrogen gas atmosphere. The QDs were purified following the previous method reported earlier.[3]

## Synthesis of $MoS_2$-CdSe and $MoS_2$-V-CdSe composites

The hydrothermal method was used to carry out the preparation of $MoS_2$-CdSe and $MoS_2$-V-CdSe QDs composites. First, 1 mmol (242 mg) $Na_2MoO_4.2H_2O$ and 5 mmol (380 mg) thiourea were dissolved in 60 ml (30 ml ethanol + 30 ml DI water) under continuous stirring for 30 minutes at room temperature to form homogeneous solutions. Then, 5 ml of freshly prepared CdSe QDs (or CdSe/$V_2O_5$ core/shell QDs) were added to the above clear solution and sonicated for 1 hour. Finally, the solution was put in a stainless-steel autoclave at 210 °C for 24 hours. After cooling to room temperature, the sample was washed several times with DI water and ethanol to remove impurities. The sample was dried in a vacuum oven at 80 °C overnight to obtain the final sample in powder form.

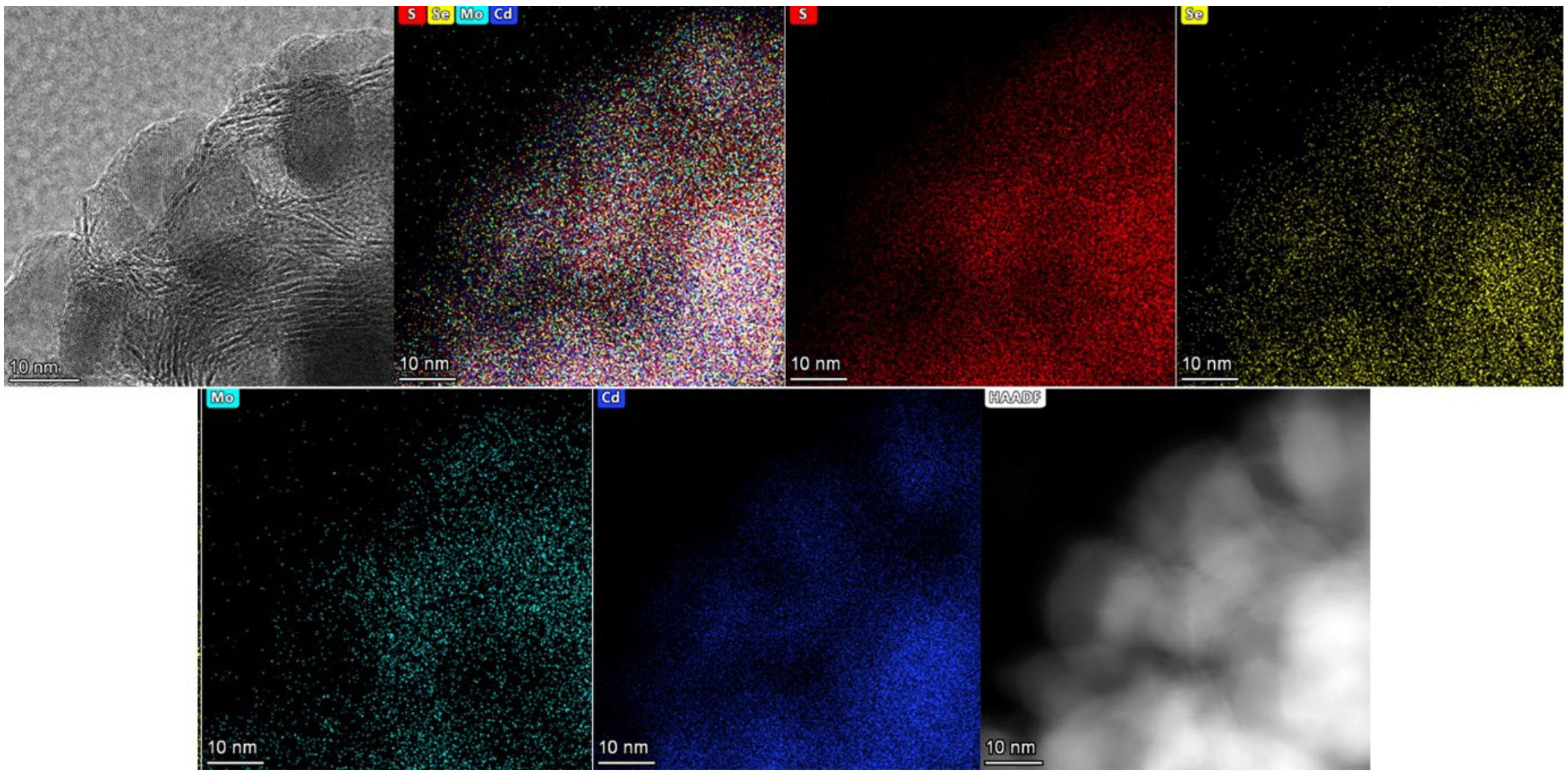

Figure S1 HRTEM image, lattice fringe image, high-angle annular dark-field (HAADF) image, and spectrum distribution with element ratio for $MoS_2$-CdSe.

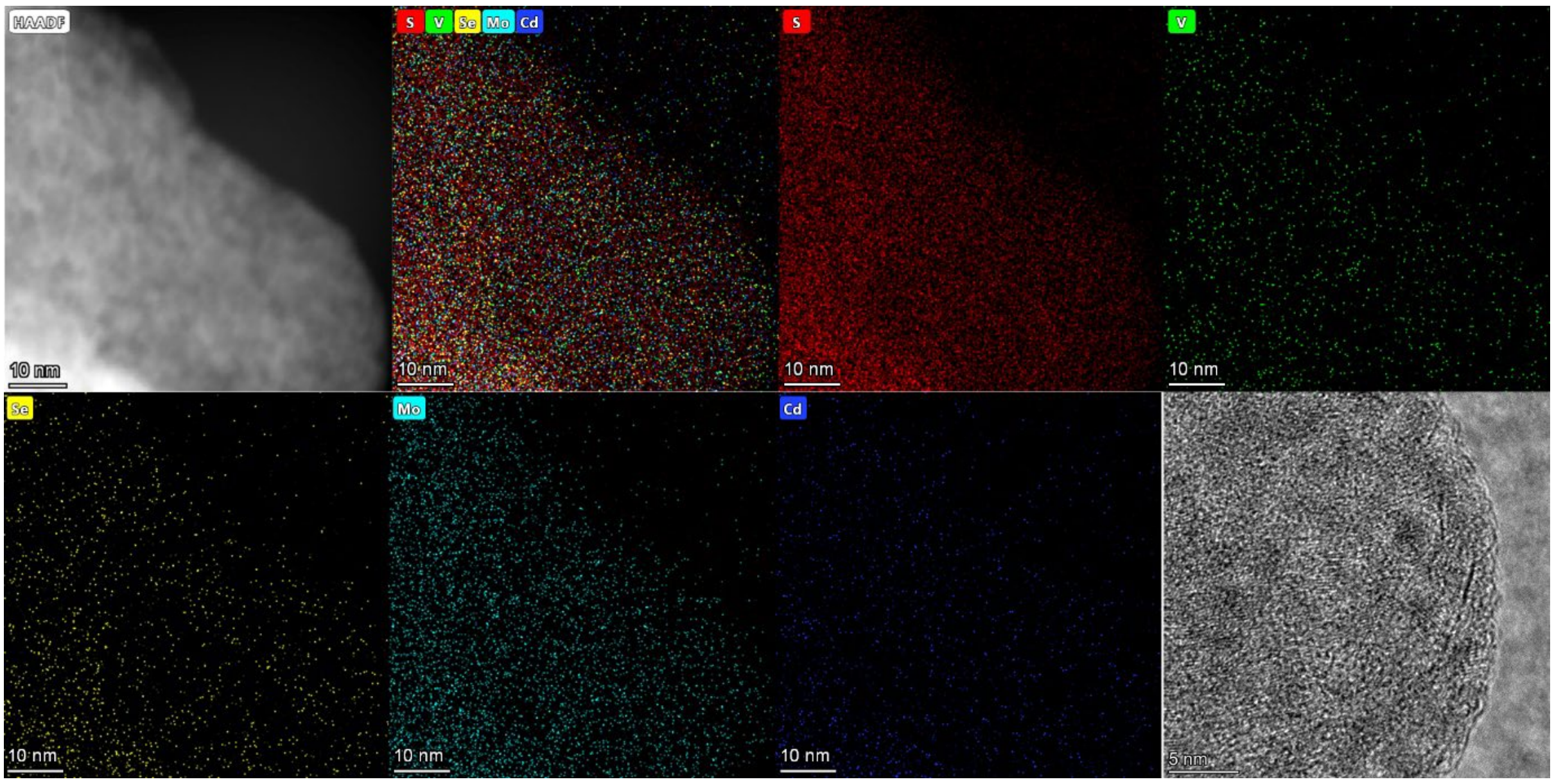

Figure S2 HRTEM image, lattice fringe image, high-angle annular dark-field (HAADF) image, and spectrum distribution with element ratio for $MoS_2$-V-CdSe.

*Table S1. Fitting parameters for samples from TRPL data.*

| Sample | TRPL Fitting parameters | |
|---|---|---|
| | $\tau_1 (ns)$ | $\tau_2\ (ns)$ |
| $MoS_2$ | 3.2614 | - |
| $MoS_2$-CdSe | 2.5608 | 7.2966 |
| $MoS_2$-V-CdSe | 2.6525 | 7.3926 |

## Sample preparation for Z-scan

The powdered sample was dispersed in DI water with a concentration of 0.1 mg/mL, followed by sonication for 30 minutes to obtain homogeneous dispersion. After centrifugation, the dispersions from the top two-thirds of the portion were gently extracted and filled in a 1 mm quartz cuvette positioned on the translation stage to move along the Z-scan path.

## Z-scans technique

The schematic and laser parameters used for Z-scan measurements have been adapted from our earlier work.[4] In brief, a Ti: Sapphire laser (Spitfire Ace, Spectra-Physics) delivers laser pulses with a pulse width of 35 fs at 800 nm with a repetition rate of 1 kHz. We used 0.2 mm β-BBO Type-I crystal to generate the second harmonic (400 nm). These 800 nm and 400 nm laser pulses were employed to obtain samples' third-order NLO properties. A spherical lens of 400 mm focal length was used to focus the laser pulses ($1/e^2$ beam radius $w_0$ =38.5 μm). For the estimation of third-order nonlinear coefficients such as absorption coefficients ($\beta$), saturation intensities ($I_{sat}$), and nonlinear refractive indices ($\gamma$) for open-aperture (OA) and closed-aperture (CA) Z-scan measurements, we used four different energies 100, 200, 280, 400 nJ (the equivalent peak intensities, $I_{800\text{ nm}}$ were 67, 135, 188 and 269 GW/cm$^2$, respectively). In contrast, for 400 nm pulses, we used 30, 50, 100, and 280 nJ pulse energies (the equivalent peak intensities, $I_{400\text{ nm}}$ were 20, 33, 67, and 188 GW/cm$^2$, respectively).

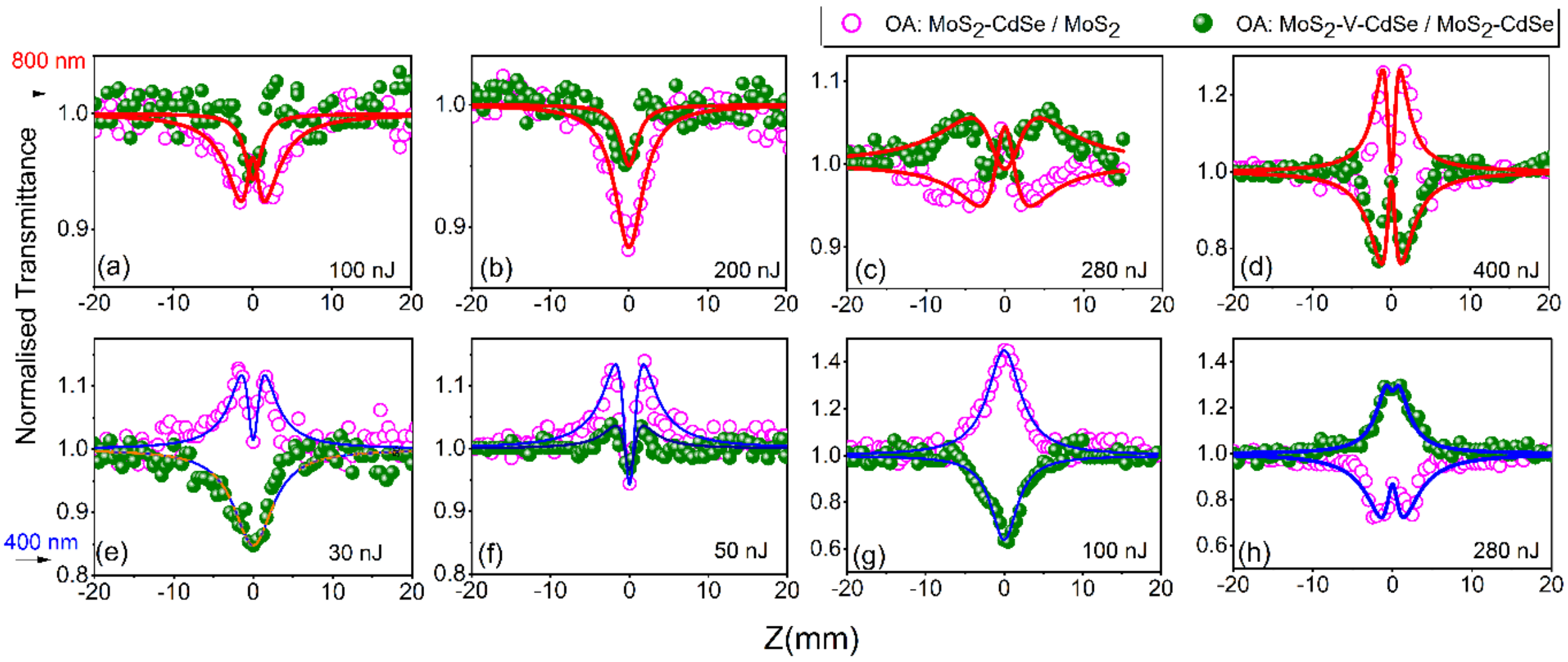

**Figure S3** *Open aperture the ratio curves of $MoS_2$-CdSe/ $MoS_2$, $MOS_2$-V-CdSe/ $MoS_2$-CdSe for (a-d) 800 nm and (e-h) 400 nm at four different pumping energies. The symbols and solid lines represent experimental data and theoretical fits, respectively.*

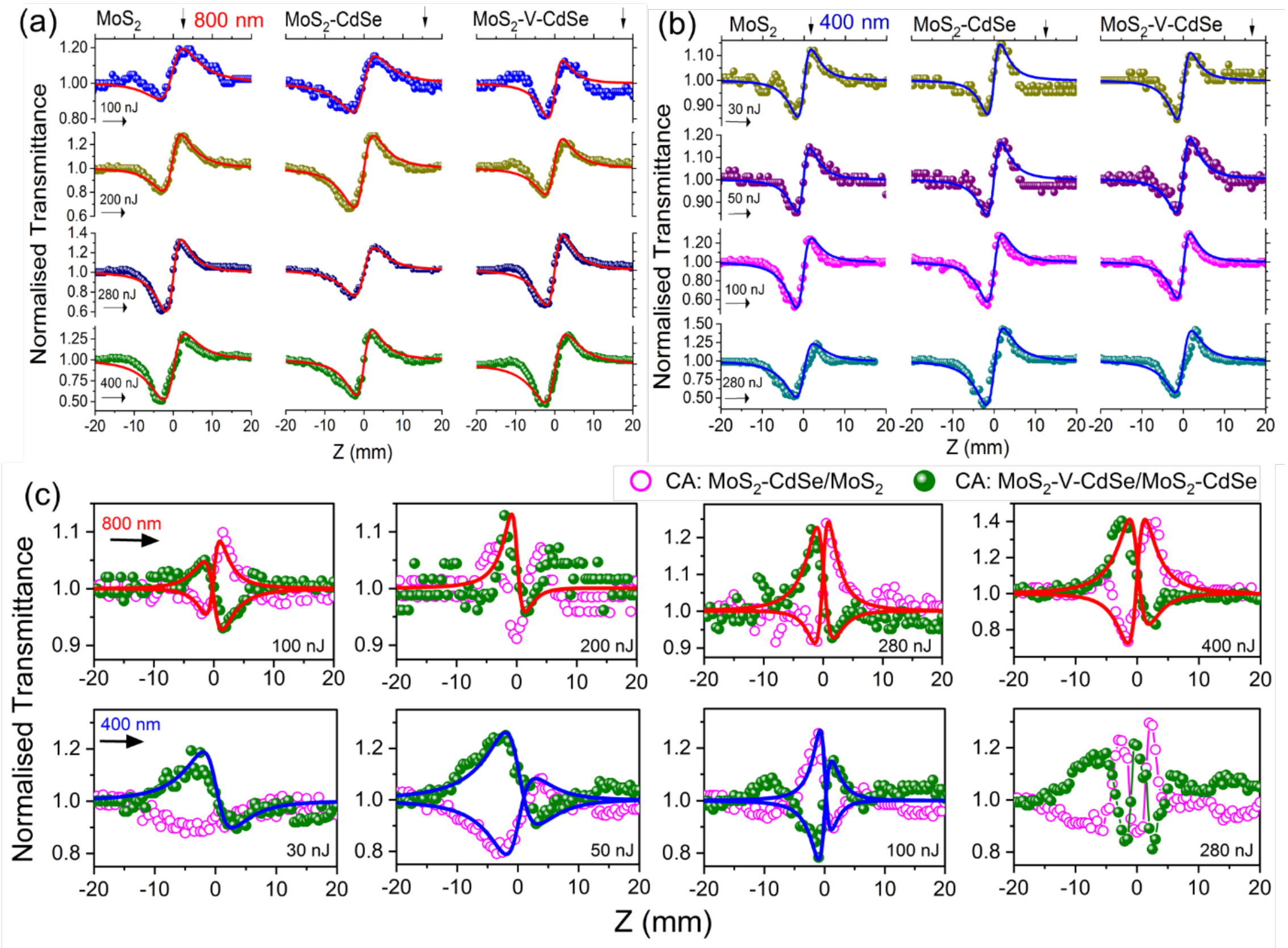

**Figure S4** *Closed aperture (CA) Z-scans result at (a) 800 nm, and (b) 400 nm for three samples. (c) the ratio curves of MoS$_2$-CdSe/ MoS$_2$, MOS$_2$-V-CdSe/ MoS$_2$-CdSe for 800 nm (top panel) and 400 nm (bottom panel) at four different pumping energies. The symbols and solid lines represent experimental data and theoretical fits, respectively.*

Table S2. Summary of NLO parameters

| **λ (nm)** | **E (nJ)** | ***I* (GW/ $cm^2$)** | **Sample** | **NLO process** | **$\beta \times 10^{-11}$ (cm/W)** | **$I_{Sat}$ ($GW/cm^2$)** | **$\gamma \times 10^{-15}$ ($cm^2/W$)** | **$Re[\chi^{(3)}]$ $\times 10^{-10}$ (esu)** | **$Im[\chi^{(3)}]$ $\times 10^{-12}$ (esu)** | **$\chi^{(3)}$ $\times 10^{-10}$ (esu)** | **FOM ($\gamma/\lambda\beta$)** |
|---|---|---|---|---|---|---|---|---|---|---|---|
| **800** | 100 | 67 | $MoS_2$ | SA | - | | 1.28 | -- | -- | -- | -- |
| | | | $MoS_2$-CdSe | - | - | | 1.46 | -- | -- | -- | -- |
| | | | $MoS_2$-V-CdSe | TPA | 1.79 | | 1.48 | 6.65 | 0.51 | 6.65 | 1.03 |
| | 200 | 135 | $MoS_2$ | SA | 1.72 | 468 | 1.09 | 4.89 | 0.49 | 4.89 | 0.79 |
| | | | $MoS_2$-CdSe | SA+TPA | 2.88 | 441 | 1.39 | 6.27 | 0.82 | 6.27 | 0.60 |
| | | | $MoS_2$-V-CdSe | SA+TPA | 7.57 | 59 | 1.07 | 4.79 | 2.16 | 4.79 | 0.17 |
| | 280 | 188 | $MoS_2$ | SA+TPA | 6.34 | 43 | 1.21 | 5.44 | 1.81 | 5.44 | 0.23 |
| | | | $MoS_2$-CdSe | TPA | 2.54 | -- | 0.86 | 3.85 | 0.72 | 3.85 | 0.42 |
| | | | $MoS_2$-V-CdSe | SA+TPA | 3.90 | 238 | 1.25 | 5.63 | 1.11 | 5.63 | 0.40 |
| | 400 | 269 | $MoS_2$ | SA+TPA | 3.35 | 303 | 0.91 | 4.07 | 0.96 | 4.07 | 0.33 |
| | | | $MoS_2$-CdSe | TPA | 4.73 | - | 0.90 | 4.05 | 1.35 | 4.05 | 0.23 |
| | | | $MoS_2$-V-CdSe | SA+TPA | 3.38 | 311 | 0.94 | 4.22 | 0.96 | 4.22 | 0.34 |
| **400** | 30 | 20 | $MoS_2$ | SA | -- | 318 | 2.08 | -- | -- | -- | -- |
| | | | $MoS_2$-CdSe | SA+ESA | 56.23 | 4 | 2.17 | 9.74 | 16.05 | 9.74 | 0.09 |
| | | | $MoS_2$-V-CdSe | SA+ESA | 31.32 | 27 | 2.05 | 9.18 | 8.94 | 9.18 | 0.16 |
| | 50 | 33 | $MoS_2$ | SA | -- | 297 | 1.32 | -- | -- | -- | -- |
| | | | $MoS_2$-CdSe | SA+ESA | 23.04 | 25 | 1.49 | 6.71 | 6.58 | 6.71 | 0.16 |
| | | | $MoS_2$-V-CdSe | SA+ESA | 25.03 | 23 | 1.52 | 6.84 | 7.15 | 6.84 | 0.15 |
| | 100 | 67 | $MoS_2$ | RSA | 13.23 | | 1.68 | 7.55 | 3.78 | 7.55 | 0.31 |
| | | | $MoS_2$-CdSe | RSA | 17.86 | | 1.65 | 7.43 | 5.09 | 7.43 | 0.23 |
| | | | $MoS_2$-V-CdSe | SA+ESA | 13.01 | 70 | 1.58 | 7.13 | 3.71 | 7.13 | 0.30 |
| | 280 | 188 | $MoS_2$ | RSA | 7.98 | | 0.59 | 2.63 | 2.28 | 2.63 | 0.18 |
| | | | $MoS_2$-CdSe | RSA | 8.93 | | 0.86 | 3.84 | 2.55 | 3.84 | 0.24 |
| | | | $MoS_2$-V-CdSe | RSA | 7.04 | | 0.69 | 3.11 | 2.01 | 3.11 | 0.24 |

Table S3. Summary of NLO parameters for ratios data.

| **λ (nm)** | **E (nJ)** | ***I* (GW/cm²)** | **Samples ratio** | **NLO process** | | **$\beta \times 10^{-11}$ (cm/W)** | **$I_{Sat}$ (GW/cm²)** | **$\gamma \times 10^{-15}$ (cm²/W)** |
|---|---|---|---|---|---|---|---|---|
| | | | | NA | NR | | | |
| **800** | 100 | 67 | $MoS_2$-CdSe / $MoS_2$ | SA+TPA | SF | 8.30 | 114 | 0.59 |
| | | | $MoS_2$-V-CdSe / $MoS_2$-CdSe | TPA | SD | 2.39 | - | -0.55 |
| | 200 | 135 | $MoS_2$-CdSe / $MoS_2$ | TPA | - | 2.45 | - | - |
| | | | $MoS_2$-V-CdSe / $MoS_2$-CdSe | TPA | SD | 1.04 | - | -0.38 |
| | 280 | 188 | $MoS_2$-CdSe / $MoS_2$ | SA+TPA | SF | 1.88 | 476 | 0.53 |
| | | | $MoS_2$-V-CdSe / $MoS_2$-CdSe | SA+TPA | SD | 2.87 | 309 | -0.47 |
| | 400 | 269 | $MoS_2$-CdSe / $MoS_2$ | SA+TPA | SF | 3.29 | 159 | 0.79 |
| | | | $MoS_2$-V-CdSe/ $MoS_2$-CdSe | SA+TPA | SD | 3.16 | 187 | -0.66 |
| **400** | 30 | 20 | $MoS_2$-CdSe / $MoS_2$ | SA+TPA | - | 33.22 | 22 | - |
| | | | $MoS_2$-V-CdSe / $MoS_2$-CdSe | SA+TPA | SD | 21.29 | - | -2.22 |
| | 50 | 33 | $MoS_2$-CdSe / $MoS_2$ | SA+TPA | SF | 23.61 | 29 | 1.33 |
| | | | $MoS_2$-V-CdSe / $MoS_2$-CdSe | SA+TPA | SD | 21.25 | 35 | -1.59 |
| | 100 | 67 | $MoS_2$-CdSe / $MoS_2$ | SA+TPA | SD | - | 149 | -0.86 |
| | | | $MoS_2$-V-CdSe / $MoS_2$-CdSe | SA+TPA | SF | 15.17 | - | 0.86 |
| | 280 | 188 | $MoS_2$-CdSe / $MoS_2$ | SA+TPA | - | 5.01 | 117 | - |
| | | | $MoS_2$-V-CdSe / $MoS_2$-CdSe | SA+TPA | - | 3.09 | 162 | - |